\documentclass[twocolumn,amsmath,amssymb]{revtex4}
\usepackage{setspace}
\usepackage{subcaption}
\usepackage{graphicx}

\usepackage{amssymb}
\usepackage{amsmath}
\usepackage[figuresright]{rotating}
\begin{document}

\title{Pressure induced antiferromagnetic-tetragonal to nonmagnetic-collapse-tetragonal insulator-metal transition in ThMnAsN}


\author{Smritijit Sen$^{1}$, Houria Kabbour$^1$ and Haranath Ghosh$^{2, 3 *}$}

\affiliation{$^1$Univ. Lille, CNRS, Centrale Lille, ENSCL, Univ. Artois, UMR 8181 – UCCS – Unité de Catalyse et Chimie du Solide, Lille, 59655 - Villeneuve d'Ascq Cedex, France.\\
$^{2}$ Theoretical and Computational Physics Section, Raja Ramanna Centre for Advanced Technology, Indore 452013, India \\
 {$^3$ Homi Bhaba National Institute, BARC Training school complex, Anushakti Nagar, Mumbai 400094, India.} \\
* corresponding author: hng@rrcat.gov.in; hngenator@gmail.com}

\begin{abstract}
We report first principles numerical discovery of hydrostatic pressure driven tetragonal to collapsed tetragonal transition in 1111-type material ThMnAsN accompanied by simultaneous magneto-structural, insulator to metal transition together with complete collapse of Mn moment. We present detailed evolution of various structural parameters, magnetism and electronic structures of ThMnAsN with increasing hydrostatic pressure. All the structural parameters show anomalies at a critical pressure P$_c \sim$ 9 GPa; c-lattice parameter, out of plane As-As bond length, anion height  (h$_{As}$) undergo drastic modification compared to the in-plane parameters which is manifested in an iso-structural phase transition from tetragonal to a collapsed tetragonal (cT) phase. In particular, above P$_c$ the out of plane As-As bond length reduces by $\sim 9\%$ whereas the h$_{As}$ shortens by $\sim 14\%$; the tetrahedral angles $\alpha / \beta $ increases/reduces by $\sim 6.4\%$/$\sim 3.6\%$; the Mn-As bond shortens by $\sim 8\%$. These modifications in "local structural correlations" due to pressure destroys usually localized nature of Mn moments and gets completely quenched. Apart from that the elastic constant, the electronic structures also bear the finger prints of insulator-metal and magneto-structural transition at higher pressures accompanying a total collapse of magnetic moment at the vicinity of 9 GPa. The critical value of the pressure  P$_c$ at which tetragonal to collapse tetragonal phase transition occurs, remains robust with respect to the on-site Hubbard correlation (U). The dynamical stability of the compound at higher pressures (above and below the magneto-structural transition) are
 affirmed through detailed computations of phonon dispersion curves endowed with positive phonon frequency through out the Brillouin zone. The effect of magnetic spin structure on the electronic band structures are obtained through band unfolding. The electronic structure of ThMnAsN at higher pressures "orbital selectively" influences bands, band gap and closely resembles with the electronic structure of Fe-based superconductors (quasi two dimensional Fermi surfaces) with the occurrences of orbital selective Lifshitz transition.  

\vspace{0.5cm}
\end{abstract}
\maketitle
\section{INTRODUCTION}
 \par In condensed matter materials research, hydrostatic pressure can be used as a non-thermal parameter that can systematically tune the ground state properties involving magnetism, superconductivity as well as structural transitions. High pressure reduces the inter-atomic distances which modifies the interaction between the quantum states through the overlap of individual atomic/electronic states causing drastic changes in the physical, magnetic, chemical as well as structural aspects. Thus,  in contrast to the real space, the effect of hydrostatic pressure in reciprocal space would manifest through the expansion of the effective Brillouin zone. A balanced combined effect of the above two reorganizes occupation probabilities of electronic (as well as lattice) degrees of freedom; as a result the formed electronic/phonon bands under pressure would be differently dispersed. An artifact of the same is perhaps to create very high or very low density of states at the redefined Fermi level, influencing visible but unpredictable changes \cite{science2016} not only to the electronic properties but also to the magnetic and superconducting ones. 

\par A plethora of ZrCuSiAs (1111) type compounds with diverse physical properties have been synthesized; largely because of the versatile nature of ZrCuSiAs type crystal structure in which all the four crystallographic sites can be occupied by a variety of elements to make it a stable compound\cite{Pottgen2008, Quebe2000, Ishida2009, Ozawa2008}. For instance, LaFeAsO which is the parent compound of the first Fe based superconductor reported by Kamihara {\it et al.,} posses ZrCuSiAs (1111) type structure consisting of alternate stacking of [La$_2$O$_2$]$^{2+}$ and [Fe$_2$As$_2$]$^{2-}$ layers with spin stripe type anti-ferromagnetic (AFM) order \cite{Kamihara2008, Muir2012}. Superconductivity arises in these compounds either by means of external pressure or by chemical doping (chemical pressure). However, a number of similar iso-structural compounds with different physical properties have been reported. For example, LaCrAsO shows metallic anti-ferromagnetism \cite{Park2013}. On the other hand, LaMnAsO displays insulating anti-ferromagnetic behaviour (band gap $\sim$ 1.2 eV) \cite{McGuire2016}, whereas, LaCoAsO is an itinerant ferromagnet in contrast to the Pauli paramagnetism with low temperature phonon mediated superconductivity in LaNiAsO \cite{Yanagi2008, Watanabe2008, Watanabe2007}. Mn based 1111 compounds including LaMnAsO in general, exhibit insulating behaviour with AFM ordering, which is of special interest because of its resemblance with the parent compounds of high T$_c$ cuprate superconductors. Apart from that, there are a number of Mn based compounds manifesting various phase transitions like Mott transition, charge ordering, spin reorientation transition etc., with the variation of temperature, external pressure, doping and applied magnetic field \cite{Kunes2008, Corkett2015, Kawano1997, Luo2007, Kim2019, Wildman2015}. In LaMnAsO, Sr doping at La site leads to a phase transition from AFM insulator to metallic AFM phase \cite{Sun2012}.
Metal-insulator transition is also reported in LaCr$_{1-x}$Mn$_x$AsO, at x$\sim$ 0.2 \cite{Park2013}. It is also reported that, AFM order in LaMnAsO can be fully suppressed by hydrostatic pressure yet no superconductivity has been discerned up to 1.5 K \cite{Guo2013, Simonson2012}. However, there is no spin reorientation transition in LaMnAsO as observed in NdMnAsO, PrMnSbO, and CeMnAsO at low temperatures \cite{Corkett2015, Marcinkova2010}. It is also worthy to be mentioned here that pressure induced metallization on the verge of metal-insulator transition is observed in BaMn$_2$As$_2$ experimentally along with the proclaimed superconductivity \cite{Satya2012}. Further, tetragonal to collapsed tetragonal (cT) phase under hydrostatic pressure is well-known in the CaFe$_2$As$_2$ and various doped Fe-based high temperature superconductors \cite{Kreyssig2008, Stavrou2015, Long2013, Uhoya2010, Mittal2011, Nakajima2015} where the cT phase is non-magnetic and the Fe-moments get quenched questioning spin fluctuation mediated superconductivity. 

\par A new 1111-type Fe based superconductors ThFeAsN has been discovered with a superconducting T$_c$ as high as 30 K without external pressure or chemical doping \cite{Wang2016}. Although no long range magnetic order has been found in ThFeAsN, a strong magnetic fluctuation above 35 K has been revealed via muon-spin rotation/relaxation and nuclear magnetic resonance techniques \cite{Shiroka2017}. On the other hand, ThNiAsN is an electron-phonon coupled superconductor with a T$_c$ of 4.3 K as evident from the previous experimental and theoretical investigations \cite{Wang2017, Yang2018}. However, a weak ferromagnetic ordering has been predicted theoretically in ThCoAsN \cite{Sen2020PRB}. ThMnAsN has been synthesized very recently by Zhang {\it et al.}, and one common thing in all these ThZAsN (Z=Fe, Co, Ni, Mn) crystal structures is distinctively shorter c-axis in contrast to the other ZrCuSiAs (1111) type compounds which give rise to a inherent internal chemical pressure \cite{Zhang2020}. This internal in-built chemical pressure is believed to play an important role in the diverse physical properties of this series of materials. Neutron diffraction data shows an AFM ordering in ThMnAsN at room temperature (300 K). Temperature variation of experimentally measured magnetic susceptibility exhibits cusps at 52 K and 36 K respectively for ThMnAsN and ThMnPN. This susceptibility cusps are indicating a spontaneous anti-ferromagnetic to anti-ferromagnetic transition for Mn$^{2+}$ moments as reported in the reference \cite{Zhang2020}. Ordered magnetic moments of ThMnAsN and ThMnPN are 3.41 $\mu_B$ (2.30 $\mu_B$) and 3.60 $\mu_B$ (2.69 $\mu_B$) respectively, measured at 4 K (300 K) \cite{Zhang2020}. The range of values of Mn moments can be well understood in terms of the half-filling of Mn-3d orbitals, which maximizes Hund’s coupling and the reduction of moment from the full moment of gS = 5 $\mu_B$ is due to the valence fluctuations \cite{Simonson2012}. Theoretical calculations on ThMnAsN also suggests an C-AFM ground state (see Fig. 1, for the magnetic configuration) with small energy gap at the Fermi level \cite{Gu2021,Sen2023}. An extensive investigation revealing the structural parameters like bond length, bond angle and electronic structure is required to understand the physical properties of ThMnAsN. Recent experimental as well as theoretical studies on ZrCuSiAs (1111) type compound also reveal that the electronic structure is very sensitive to pressure (chemical/external) \cite{Zhang2020, Sen2020PRM, Satya2012}. Further as mentioned earlier there are a number of Fe-based high temperature superconductors that undergoes tetragonal to the cT transition \cite{Kreyssig2008, Stavrou2015, Long2013, Uhoya2010, Mittal2011, Nakajima2015}. Therefore, it is compulsive to study the effect of external pressure on the magnetism and electronic structures of ThMnAsN.
 
\par In this work, we show that as a result of hydrostatic pressure ThMnAsN undergoes a magneto structural phase transition from tetragonal to cT phase transition in which Mn moments are completely quenched; very similar to those in Fe-based superconductors \cite{Kreyssig2008, Stavrou2015, Long2013, Uhoya2010, Mittal2011, Nakajima2015} indicating a possibility of Mn based high temperature superconductivity under pressure. Further, our simulated results suggest that at higher pressures, ThMnAsN posses very similar electronic structure as that of the Fe based superconductors. Rest of the paper is organized as follows. In Section II we describe computational methodologies used in this work. This is followed by results and discussions section III which comprises  a few subsections. In subsection III-A, we describe effects of pressure on various important  structural parameters that controls magnetism and superconductivity.  This subsection also includes important associated physical properties of ThMnAsN related to elastic constants, Cauchy pressure, electronic charge density modulation, Crystal Orbital Hamilton Population (COHP) bond analysis etc. Importantly, in this section we pin point the essential difference in the mechanism of cT phase transition in ThMnAsN as compared to those observed in 122 type Fe-based superconductors. In section III-B we present associated modifications due to hydrostatic pressures below and above the magneto structural transition in electronic structures like electronic band structure, density of states, Fermi surface etc. Important feature of this subsection is that the electronic structures are converted from (small) magnetic Brillouin zone (corresponding to the supercell structure)  to the original Brillouin zone (corresponding to the primitive cell) through band unfolding which is generally missing from the literature and would be observed in the experiments (say ARPES). This subsection is ended with a brief discussions on dynamical stability of the material at ambient as well as higher pressures through presentations of phonon dispersion and phonon density of states. Finally, we present conclusions with the essence of this work in the section IV.
 
	\section{COMPUTATIONAL METHODS}
The crystal structure of ThMnAsN is tetragonal with space group symmetry $P4/nmm$ (space group number 129). ThMnAsN consists of alternating layers of MnAs and ThN. MnAs plane is very similar to the FeAs/Se plane of the Fe based superconductors where Mn atoms are in the same plane but the As atoms are situated above and below the Mn plane. Height of the As atoms from the Mn plane is known as 'anion height'. Experimental lattice parameters of the tetragonal ThMnAsN at 4 K, are used as input of  first principles density functional theory calculations \cite{Zhang2020}. The first principles computations are carried out by employing the projector augmented-wave (PAW) method as implemented in the Vienna \textit{ab initio} simulation package (VASP) \cite{Kresse1993,Bloch1994,Kresse1996}. Exchange correlation functional has been treated under the generalized-gradient approximation (GGA) within Perdew-Burke-Ernzerhof (PBE) functional \cite{Perdew1996}.
\begin{figure}
\includegraphics[height=8cm,width=7.5cm]{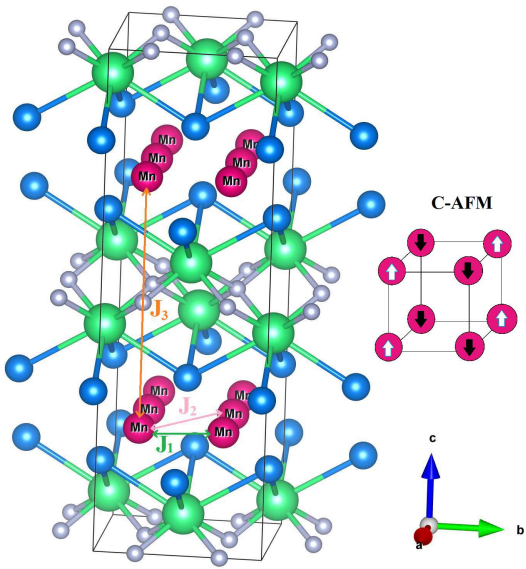}
\caption{ A schematic representation of $\sqrt{2}\times\sqrt{2}\times2$ supercell of tetragonal ThMnAsN. Th, Mn, As and N atoms are indicated by light-green, pink, blue and grey colours respectively. Various Heisenberg type magnetic couplings (J$_1$, J$_2$ and J$_3$) between Mn atoms are also shown. The c-type antiferromagnetic (C-AFM) spin configuration is presented in the right hand side of the figure.} 
\label{SC}
\end{figure}
We construct a $\sqrt{2}\times\sqrt{2}\times2$ magnetic super-cell in order to implement the C-AFM spin arrangements. A schematic diagram of the $\sqrt{2}\times\sqrt{2}\times2$ super-cell of ThMnAsN is shown in Fig.\ref{SC}. Various magnetic coupling constants (J$_1$, J$_2$, J$_3$) are also indicated in the Fig.\ref{SC}.
We perform spin polarised calculations using C-AFM spin configuration. The wave functions were expanded in plane wave basis with an energy cutoff of $550$ eV and the energy tolerance of the self consistent calculations are set to 10$^{-6}$ eV. The sampling of the Brillouin zone is performed using a $\Gamma$-centered $6\times6\times3$ Monkhorst-Pack grid.
To obtain the crystal structures at different pressures, we start with P=0 structure. Then, we increase the pressure gradually at the steps of 1 GPa and relax the lattice parameters and internal atomic coordinates at each pressure. Density of states and band structure calculations are performed using these optimized crystal structures at a particular external pressure. The forces and stresses of the converged structures at each pressure were optimized and checked to be within the error allowance of the VASP code. Elastic constants were calculated within VASP by finite differences of stress with respect to strain \cite{Page2002}. For the Fermi surface calculations in the non magnetic phase, a denser k-grid of size $20\times20\times20$ is considered for the calculation. In order to visualise the Fermi surface, we use Fermi surfer software \cite{Kawamura2019}. Charge density plots are prepared using VESTA software \cite{Momma2008}. The COHP analysis are carried out in the framework of the LOBSTER software \cite{Maintz2016, Deringer2011, Dronskowski1993}. Phonon calculation has been carried out within density functional perturbation theory using VASP and PHONOPY \cite{Togo2015} codes.
	\section{RESULTS AND DISCUSSION}
 In this section, we will demonstrate the effect of hydrostatic pressure in the structural parameters, magnetism and electronic structures of ThMnAsN. In the first step, we fully optimize (both lattice parameters as well as atomic positions) the geometry of the crystal structure. A number of magnetic configurations (spin arrangement of Mn) are considered. The C-AFM as shown in Fig.1 turns out to be the ground state configuration.  In Table-\ref{table2}, we present the GGA optimized structural parameters as well as the local magnetic moment of Mn atom for ThMnAsN with C-AFM spin configuration (see Fig. 1). Along with that, we also depict the experimentally measured structural parameters and Mn moment at 4 K in Table-\ref{table2} (last column). Changes in the optimized lattice parameters with the experimental values are less than 0.2 $\%$ whereas 1.4 $\%$ decrease in the z$_{As}$ value is observed. This decrement of GGA optimized z$_{As}$ from the experimental value is quite similar to the case of Fe based superconductors and it is an indication of the presence of magnetic fluctuations in the system \cite{Mazin2008}. On the other hand, local magnetic moment of Mn in the optimized structure is in good agreement with the experimentally measured value. GGA+U, HSE06 calculations and the effect of spin-orbit coupling on the C-AFM ground state of ThMnAsN are discussed in the supplementary materials (SM).
 \begin{figure}[h!]
		\centering
		\includegraphics [height=7cm,width=8.5cm]{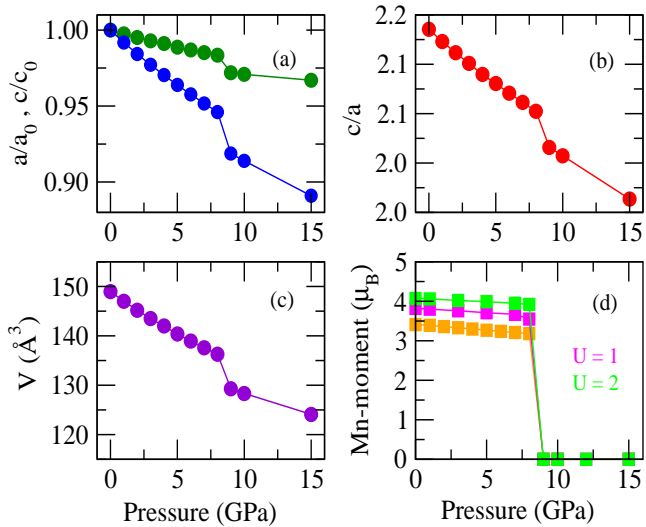}
		\caption{Variation of (a) Normalized lattice parameters (a/a$_0$, c/c$_0$), (b) c/a ratio, (c) volume of the unit cell (V) and (d) local magnetic moment (Hubbard U = 1 eV and 2 eV on Mn-d orbital) of Mn atom with hydrostatic pressure for ThMnAsN with C-AFM spin configuration.}
		\label{Pstr}
	\end{figure}
 \begin{figure}[h!]
		\centering
		\includegraphics [height=7cm,width=8.5cm]{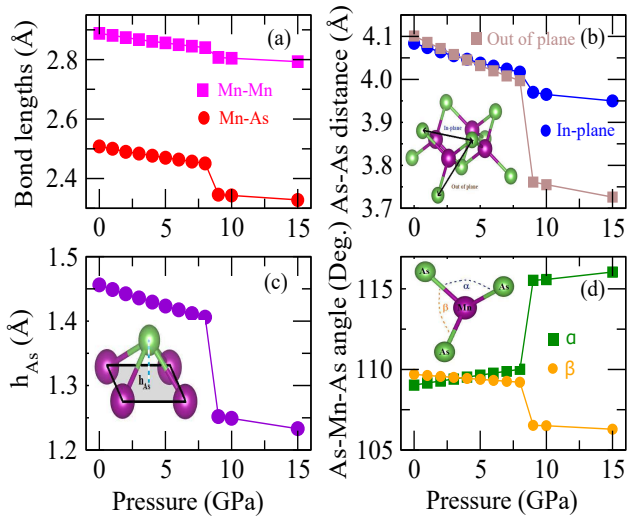}
		\caption{Variation of (a) bond lengths (Mn-Mn, Mn-As), (b) In-plane and out of plane As-As distances, (c) anion height (h$_{As}$) and (d) As-Mn-As bond angles ($\alpha$, $\beta$) with hydrostatic pressure for ThMnAsN with C-AFM spin configuration.}
		\label{Pbond}
	\end{figure}

\begin{figure}[h!]
		\centering
		\includegraphics [height=4cm,width=8.8cm]{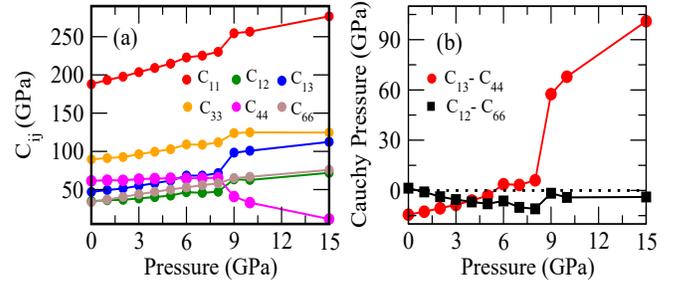}
		\caption{Pressure variation of (a) various elastic constants and (b) Cauchy pressures (C$_{13}$-C$_{44}$) and (C$_{12}$-C$_{66}$) of tetragonal ThMnAsN.}
		\label{elastic}
	\end{figure}
 \begin{figure}[h!]
		\centering
		\includegraphics [height=5.5cm,width=6.5cm]{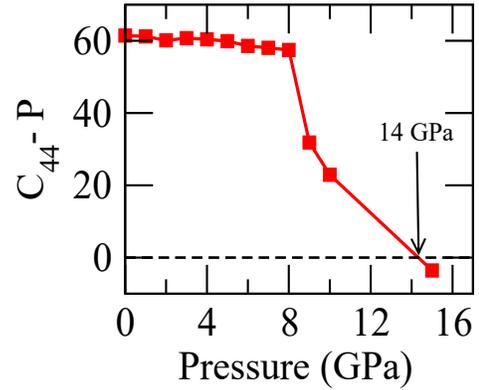}
		\caption{Variation of C$_{44}$-P as a function of external pressure for tetragonal ThMnAsN.}
		\label{c44}
	\end{figure}
 
	\begin{table}[h!]
		\begin{center}
			\caption{Optimised and experimental (measured at 4 K) structural parameters and local magnetic moment of Mn for ThMnAsN with the C-AFM spin arrangement.}
			\vspace{2mm}
			\label{table2}
			\begin{tabular}{l|c|c} 
				\hline
				{Structural} & GGA & Experiment*\\
				{parameters} & (C-AFM) & (4 K) \\
				\hline
				a=b (\AA) & 4.084 & 4.076 \\
				\hline
				c (\AA) & 8.927 & 8.929 \\
				\hline
				V (\AA$^3$) & 148.89 & 148.34 \\
				\hline
				z$_{Th}$  & 0.1293 & 0.1301 \\
				\hline
				z$_{As}$  & 0.6631 & 0.6727 \\
				\hline
				Mn moment ($\mu_B$) & 3.42  & 3.41 \\
				\hline
			\end{tabular}
		\end{center}
		\begin{center}
		* Reference \cite{Zhang2020}
		\end{center}
	\end{table}
	\subsection{Effect of hydrostatic pressure on the structural parameters and magnetism of ThMnAsN}
    
    In the context of high temperature superconductivity in Fe-based superconductors, correlation of local structural parameters like h$_{As}$, As-Fe-As angles etc.  with T$_c$ (superconducting transition temperature) and its effect in electron correlation and magnetism is well established \cite{Chen2014, Mizuguchi2010, MingY2017}. In Fig.\ref{Pstr}, we display the variation of various structural parameters like normalized lattice parameters (a/a$_0$, c/c$_0$), c/a ratio, volume of the unit cell (V) and local magnetic moment of Mn atom as a function of hydrostatic pressure. It is quite evident that there is an abrupt change in the normalized lattice parameters, c/a ratio and volume of the unit cell at 9 GPa pressure. The change in the c-axis as compared to the a-axis is more prominent. This anomalous behaviour of structural parameters with the hydrostatic pressure is attributed to a iso-structural phase transition (tetragonal to collapse tetragonal) that occurs simultaneously with the magnetic transition (AFM to PM phase). It is worth mentioning here that in LaMnPO, a iso-structural compound to ThMnAsN, pressure induced AFM to PM transition is observed experimentally at 34 GPa \cite{Guo2013}. Furthermore, we also study the variation of various bond lengths and bond angles with pressure as shown in Fig.\ref{Pbond}. From Fig.\ref{Pbond}a, we can see a pronounced change in the Mn-As bond distance at 9 GPa pressure as compared to the Mn-Mn bond distance. Added to that, Fig.\ref{Pbond}b(c), showing the variation of in plane and out of plane As-As distances ('anion height') with pressure also indicate a significant out of plane modification (along the c-axis) that occurs at 9 GPa. Modifications of Mn-As tetrahedral bond angles ($\alpha$ and $\beta$) are depicted in Fig.\ref{Pbond}d. It is quite evident from Fig.\ref{Pbond}d that, a large distortion in the Mn-As tetrahedron transpire at 9 GPa pressure. 
 \begin{figure}[h!]
		\centering
		\includegraphics [height=7.5cm,width=7.5cm]{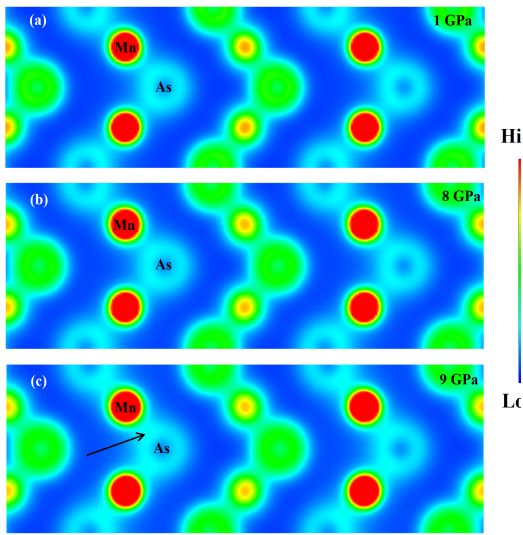}
		\caption{Calculated charge density plots of ThMnAsN along (100) plane at (a) 1 GPa (b) 8 GPa and (c) 9 GPa of hydrostatic pressure.}
		\label{Chrgp}
	\end{figure}
 \begin{figure}[h!]
	\centering
\includegraphics [height=7.5cm,width=8.5cm]{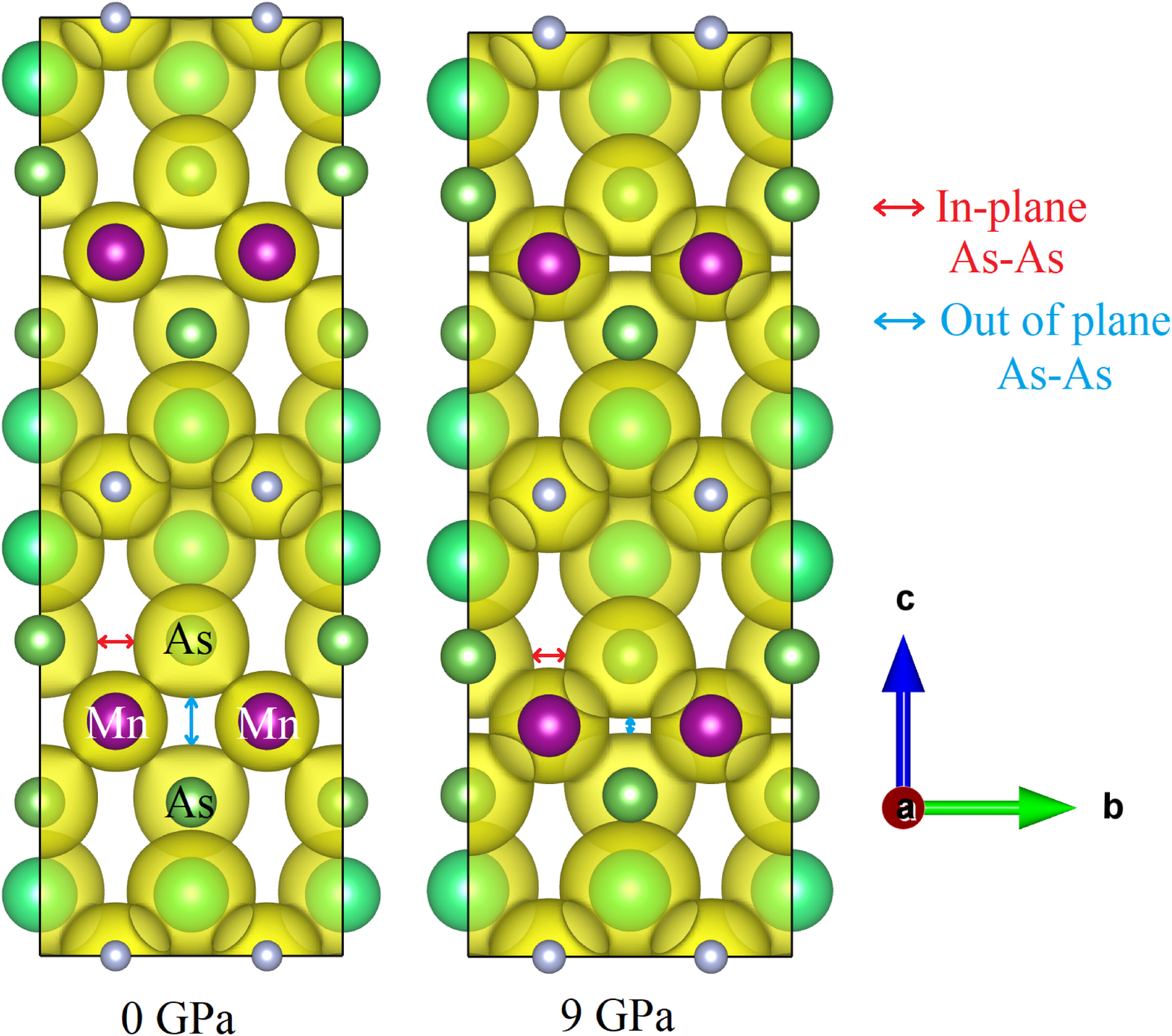}
\caption{Calculated isocharge surface of electron density for ThMnAsN at ambient and 9 GPa of hydrostatic pressure.}
		\label{isop}
	\end{figure}
	\begin{figure}[h!]
		\centering
		\includegraphics [height=5cm,width=8.5cm]{LOBS.eps}
		\caption{COHP bonding analysis of Mn-As bonds (red for spin up and black for spin down) at (a) 1 GPa, (b) 8 GPa, (c, d) 9 GPa pressure (C-AFM and NM).}
		\label{LOBS}
	\end{figure}
 Thus, at around 9 GPa pressure, all the structural parameters show strong anomalous variations e.g, c-lattice parameter, out of plane As-As bond length, anion height  (h$_{As}$) undergo drastic modification. In particular, above P$_c$ the out of plane As-As bond length contracts by ~ 9 $\%$ whereas the h$_{As}$ shortens by $~ 14 \%$; the tetrahedral angles $\alpha / \beta $ increases/decreases by 
  ~ 6.4 $\% /$ ~3.6 $\%$; the in-plane Mn-As bond shortens by $~ 8 \%$. The out of plane As-As bond length  becomes shorter than its planer counter part. These modifications of "local structural correlation" influences the electronic coordination of otherwise localized magnetic moments of Mn atoms which gets collapsed (quenched) at the same critical pressure (9 GPa) indicating a magnetic phase transition from AFM to para-magnetic (PM) phase. The critical value of the pressure P$_c$ = 9 Gpa is robust and does not get modified with increasing Hubbard correlation U (Mn-d) up to 5 eV. However, the value of the magnetic moment of Mn in the C-AFM phase for GGA+U calculations exceeds the experimental value (see discussion in SM) and hence not presented in the figure (for higher values of U). This indicates to the fact at higher pressures the value of the Hubbard correlation for ThMnAsN might have been reduced from that of the ambient pressure. The two different phase transitions,  magnetic and structural one occurs simultaneously driven by each other. Such simultaneous magneto-structural transition is very similar to those observed experimentally in NaFe$_2$As$_2$, CaFe$_2$As$_2$ and several other Fe-based superconductors \cite{Kreyssig2008, Stavrou2015, Long2013, Uhoya2010, Mittal2011, Nakajima2015} except the fact that the in-plane and out-of-plane As-As bond distances are comparable for ThMnAsN.  The ratio c/a of the tetragonal phase at which cT phase transition takes place is about 2.1 for ThMnAsN whereas about 2.9 for KFe$_2$As$_2$ / CaFe$_2$As$_2$ / BaFe$_2$As$_2$ compounds respectively. The volume contraction is about 4.0 $\%$ for ThMnAsN whereas is about 3.0$\%$ in  CaFe$_2$As$_2$ / BaFe$_2$As$_2$ compounds respectively.  This opens up a prospects for ThMnAsN to possibly superconduct at higher pressures (superconductivity is observed in KFe$_2$As$_2$ \cite{Nakajima2015} above the cT phase with pressure). 
\par Next we present the variation of elastic constants of tetragonal ThMnAsN as a function of hydrostatic pressure as depicted in Fig.\ref{elastic}a. Mechanical stability of a crystal under external pressure requires to satify the generalised Born-Huang stability criteria. For a tetragonal system the Born-Huang stability criteria is defined as \cite{Gomis2014}: C$_{11}$-P$>$0,
	C$_{44}$-P$>$0,	C$_{66}$-P$>$0, C$_{11}$-C$_{12}$-2P$>$0 and (C$_{33}$-P)(C$_{11}$+C$_{12}$)-2(C$_{13}$+P)$^2>$0. All the above criteria is satisfied for ThMnAsN up to 14 GPa pressure.
	Fig.\ref{elastic}a shows that all the elastic constants are more or less increasing with the increase of external pressure except C$_{44}$. It is also to be noted here that the elastic constant C$_{11}$ is always greater than C$_{33}$ up to 15 GPa pressure which indicates that it is easier to compress the system along 001 direction than in 100 direction. In order to understand the observed behaviour of elastic constants around 9 GPa pressure, we also calculate the Cauchy pressure for tetragonal system as a function of pressure (see Fig.\ref{elastic}b). It is well known that for metallic bonding, the Cauchy pressure is positive, while for directional bonding, it is negative and larger negative Cauchy pressure indicates a more directional character in the bonds \cite{Pettifor1992}. As we can see from Fig.\ref{elastic}b, Cauchy pressure C$_{13}$-C$_{44}$ has small negative value at lower pressure but it increases rapidly as we go above 9 GPa pressure. On the other hand, there is almost no variation of Cauchy pressure C$_{12}$-C$_{66}$ with the hydrostatic pressure. This is so, because the variations of elastic constants C$_{12}$ and C$_{66}$ with the pressure are not very substantial as compared to the other elastic constants where the pressure variation is noticeable. High positive values of Cauchy pressure C$_{13}$-C$_{44}$ above 9 GPa pressure indicate the enhancement of metallic behaviour. In Fig.\ref{c44}, we present the pressure variation of C$_{44}$-P. There is a sudden drop of C$_{44}$-P at around 9 GPa pressure and above 14 GPa pressure C$_{44}$-P becomes negative. When the value of C$_{44}$-P is less than zero, it indicates that the system is mechanically unstable or is on the threshold of structural phase transition \cite{Gomis2014}. 
\par Further, we have also depicted the charge densities of ThMnAsN at various pressures along the 100 plane in Fig.\ref{Chrgp}. We can clearly see that (see the black arrow in the Fig.\ref{Chrgp}) at 9 GPa pressure, the overlap of charge densities along Mn-As bond is significantly increased if we compare the same with that at the 8 GPa pressure. All these results collectively stipulate the enhancement of metallic character above 9 GPa pressure. In Fig.\ref{isop}, we depict the iso-surface of electronic charge densities at ambient as well as 9 GPa pressure. At 9 GPa the chrage densities of the out of plane As atoms are getting very close to each other, however, no direct covalent bonding is formed unlike those observed in 122 family of Fe-based superconductors \cite{Kreyssig2008, Stavrou2015, Long2013, Uhoya2010, Mittal2011, Nakajima2015}. As the out of plane As-As distances become less than the in-plane As-As bond length, all 4-As atoms strongly overlap with the Mn atom. This leads to indirect overlapping between the Mn atoms via As atoms (see Fig. \ref{isop}). This destroys the otherwise "localized" nature of Mn moments and moments get completely quenched above P$_c$. 
\par To corroborate the charge density plots at 1, 8 and 9 GPa, the calculated integral crystal orbital Hamilton population (ICOHP) is indicated on Table \ref{table3}. Their sign indicates net bonding for the reported distances. The absolute value of the ICOHP is proportional to the strength of the bond and is increasing for all bonds with pressure from 1 to 9 GPa. The Mn-As absolute value shows an increase of 6.7 \% from 1 to 8 GPa and of 8.9 \% from 8 to 9 GPa, considering the up spin channel (similar trend is observed for the dn spin channel). Therefore, the evolution in the smaller range 8 to 9 GPa is more significant than the whole 1 to 8 GPa range. It signifies the structural transition described from 8 to 9 GPa and the enhancement of metallic character at 9 GPa. On the other hand, projected COHP is shown in Fig.\ref{LOBS} for the bonds Mn-As. Positive COHP represents bonding interactions while a negative COHP represents anti-bonding interactions. Anti-bonding states are present at the top of the valence band just below the Fermi level for both 1 and 8 GPa. In these two cases [Fig.\ref{LOBS}a, b], the band gap is opened in the AFM state while the metallic state at 9 GPa gives rise to anti-bonding states at the Fermi level. Such situation is destabilizing and should lead to a structural distortion and/or magnetic transition.
	\begin{table}[t]
		\begin{center}
			\caption{ICOHP (Integral Crystal Orbital Hamilton Population) for ThMnAsN for the different types of bonds at 1, 8 and 9 GPa of hydrostatic pressure.}
			\vspace{2mm}
			\label{table3}
			\begin{tabular}{l|c|c|c|c|c|c} 
				\hline
				{Press}&Mn-As&ICOHP&Th-N &ICOHP&Th-As &ICOHP\\
    {(GPa)}& (\AA)&($\uparrow$/$\downarrow$)&(\AA)&($\uparrow$/$\downarrow$)&(\AA) &($\uparrow$/$\downarrow$)\\
				\hline
				1 & 2.508 &-3.89&	2.347&	-2.69	&3.417&	-1.47\\
    &  &/ -3.52&	&	/-2.68	&&	/-1.49\\
				\hline
				8&	2.449&	-4.15&	2.316&	 -2.80&	3.291&	-1.66 \\
    &	&/ -3.74&	&	/ -2.80&	&	/-1.68 \\
				\hline
				9 &	2.336&	-4.52&	2.309&	-2.84&	3.281&	-1.84 \\
                (AFM) & & /-4.16& &/ -2.84 & & /-1.86\\
				\hline
				9 &	2.336&	-4.37&	2.309&	-2.84&	3.281&	-1.83 \\
    (NM) & & & & & & \\
				\hline
			\end{tabular}
		\end{center}
	\end{table}
\subsection{Electronic structure of ThMnAsN at ambient and high pressure}
 \begin{figure}[h!]
\begin{subfigure}{0.50\textwidth}
   \includegraphics[height=7cm,width=10.0cm]{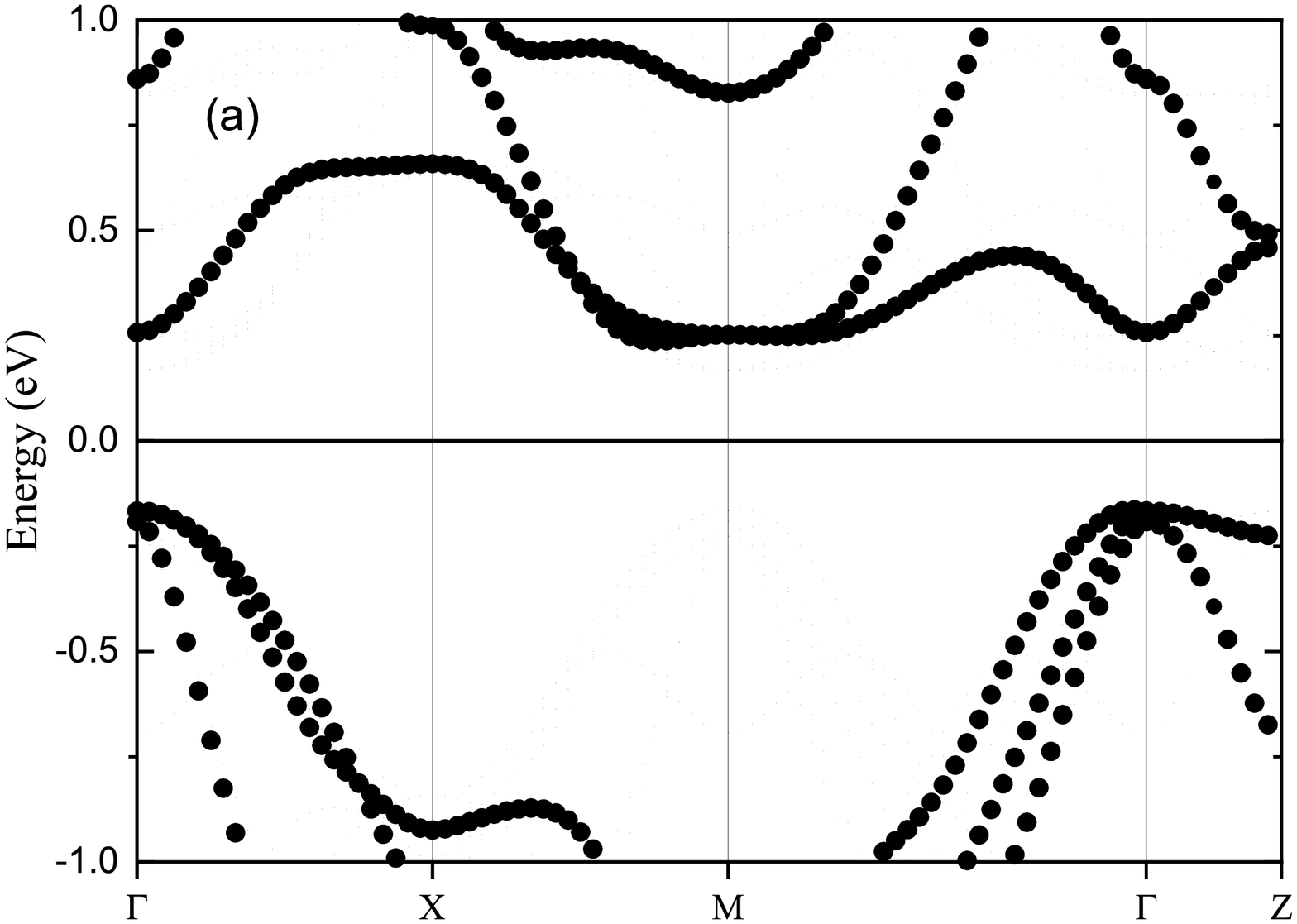}
\end{subfigure}
\begin{subfigure}{0.50\textwidth}
   \includegraphics[height=7cm,width=10.0cm]{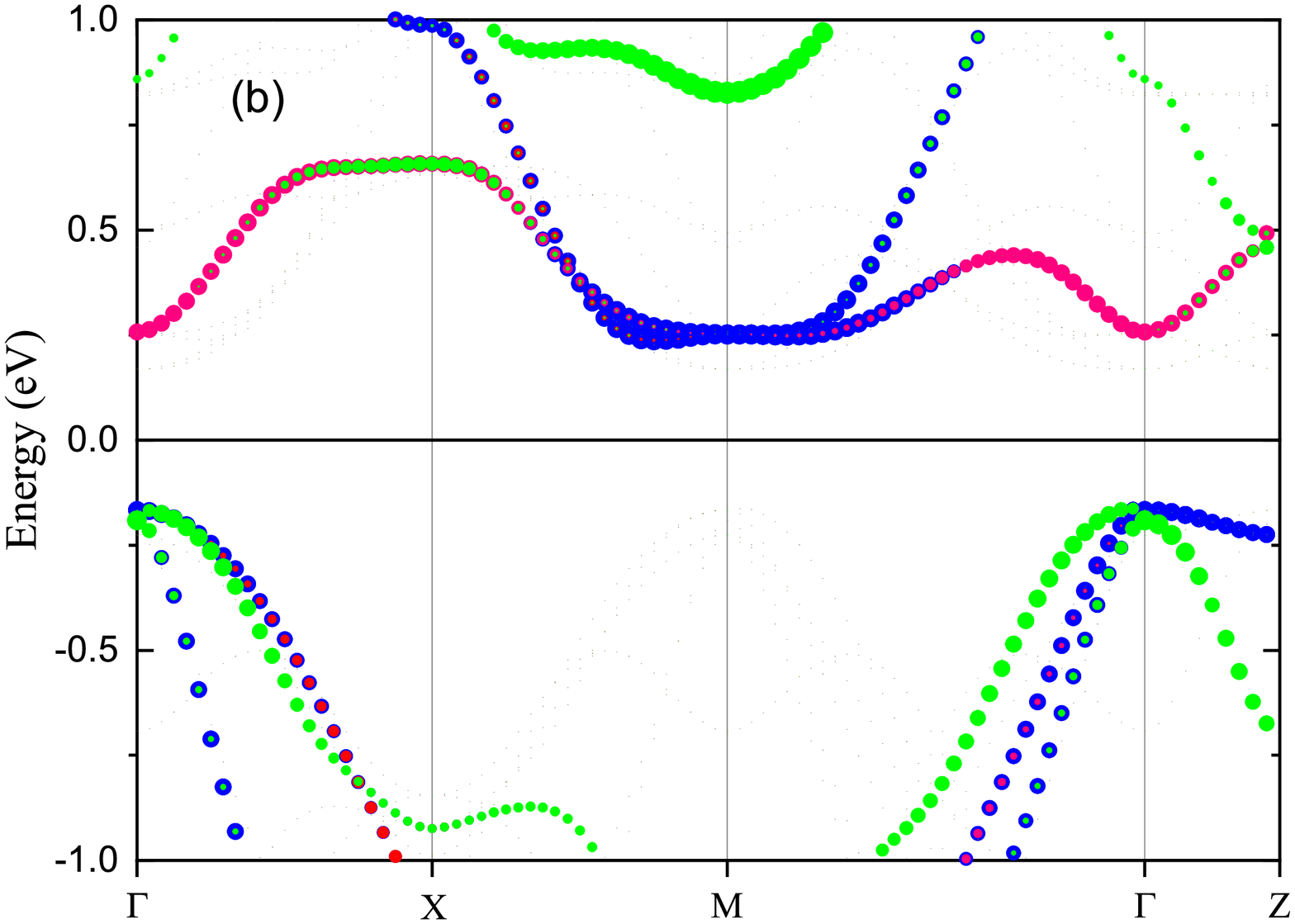}
\end{subfigure}
\caption{Calculated unfolded (a) band structure and (b) Mn-d orbital projected band structure of ThMnAsN at ambient pressure for C-AFM spin configuration. Mn d$_{xz+yz}$, d$_{xy}$, 
d$_{z^2}$ and d$_{x^2-y^2}$ orbitals are denoted by blue, red, magenta and green colours respectively.}
   \label{cafmbs}
\end{figure}

\begin{figure}
\centering
\begin{subfigure}{0.50\textwidth}
   \includegraphics[height=7cm,width=10.0cm]{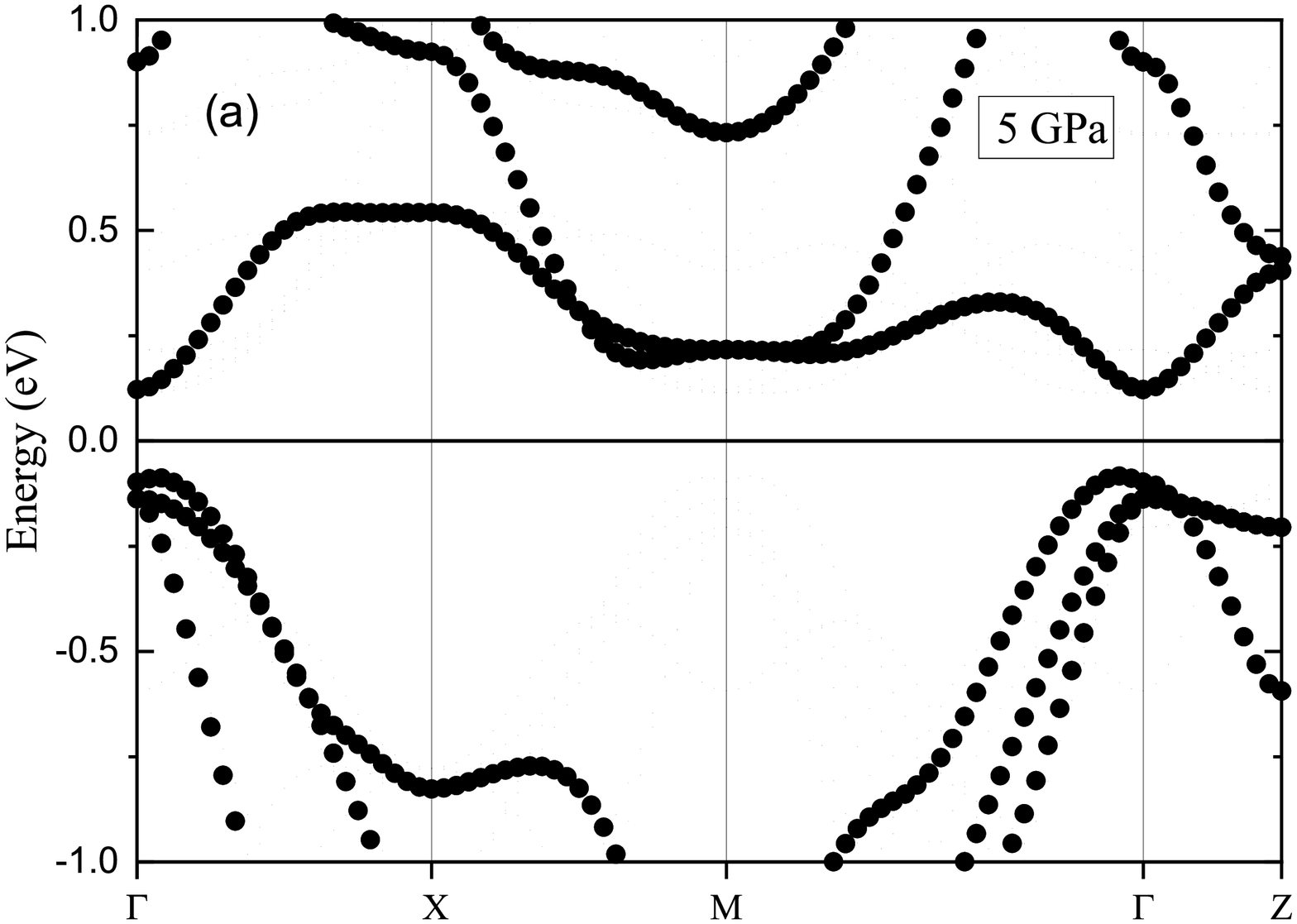}
\end{subfigure}
\begin{subfigure}{0.50\textwidth}
   \includegraphics[height=7cm,width=10.0cm]{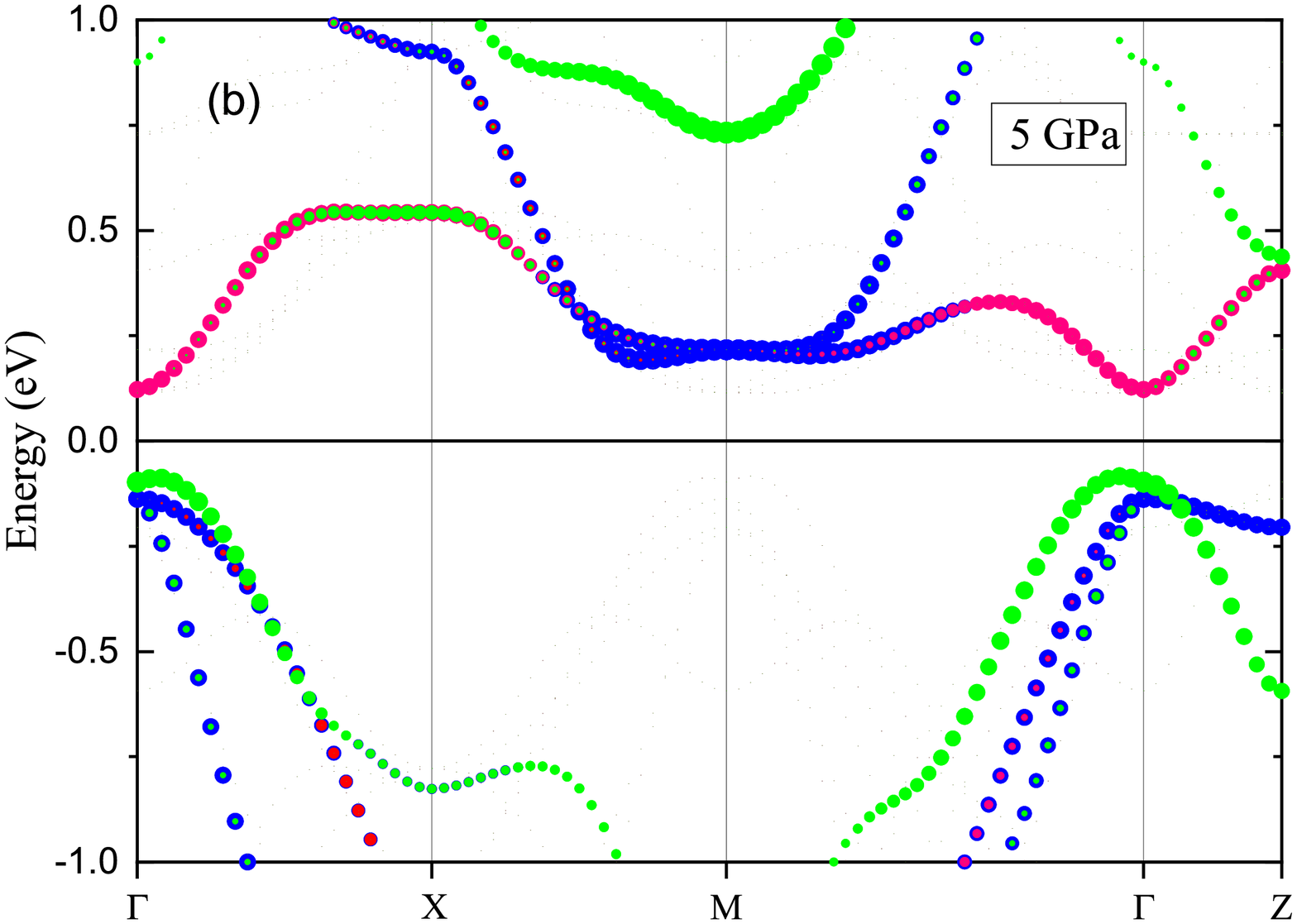}
\end{subfigure}
\caption{Calculated unfolded (a) band structure and (b) Mn-d orbital projected band structure of ThMnAsN at 5 GPa pressure for C-AFM spin configuration. Mn d$_{xz+yz}$, d$_{xy}$, d$_{z^2}$ and d$_{x^2-y^2}$ orbitals are denoted by blue, red, magenta and green colours respectively.}
   \label{5cafmbs}
\end{figure}
\begin{figure}[h!]
		\centering
		\includegraphics [height=4.5cm,width=7.0cm]{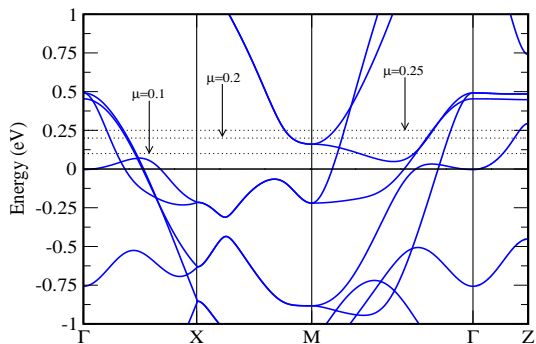}
		\caption{Calculated band structure of non-magnetic ThMnAsN at 9 GPa pressure. Black dotted lines indicate the shifted Fermi levels for various chemical potentials. }
		\label{band}
	\end{figure}
	\begin{figure}[h!]
		\centering
		\includegraphics [height=5cm,width=7.0cm]{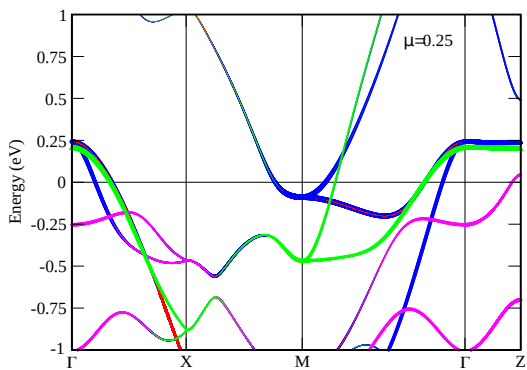}
		\caption{Calculated Mn-3d orbital projected band structure of non-magnetic ThMnAsN at 9 GPa pressure for $\mu=0.25$. Mn d$_{xz+yz}$, d$_{xy}$, d$_{z^2}$ and d$_{x^2-y^2}$ orbitals are denoted by blue, red, magenta and green colours respectively.}
		\label{orb}
	\end{figure}
	  
	\begin{figure}[h!]
		\centering
		\includegraphics [height=11cm,width=8cm]{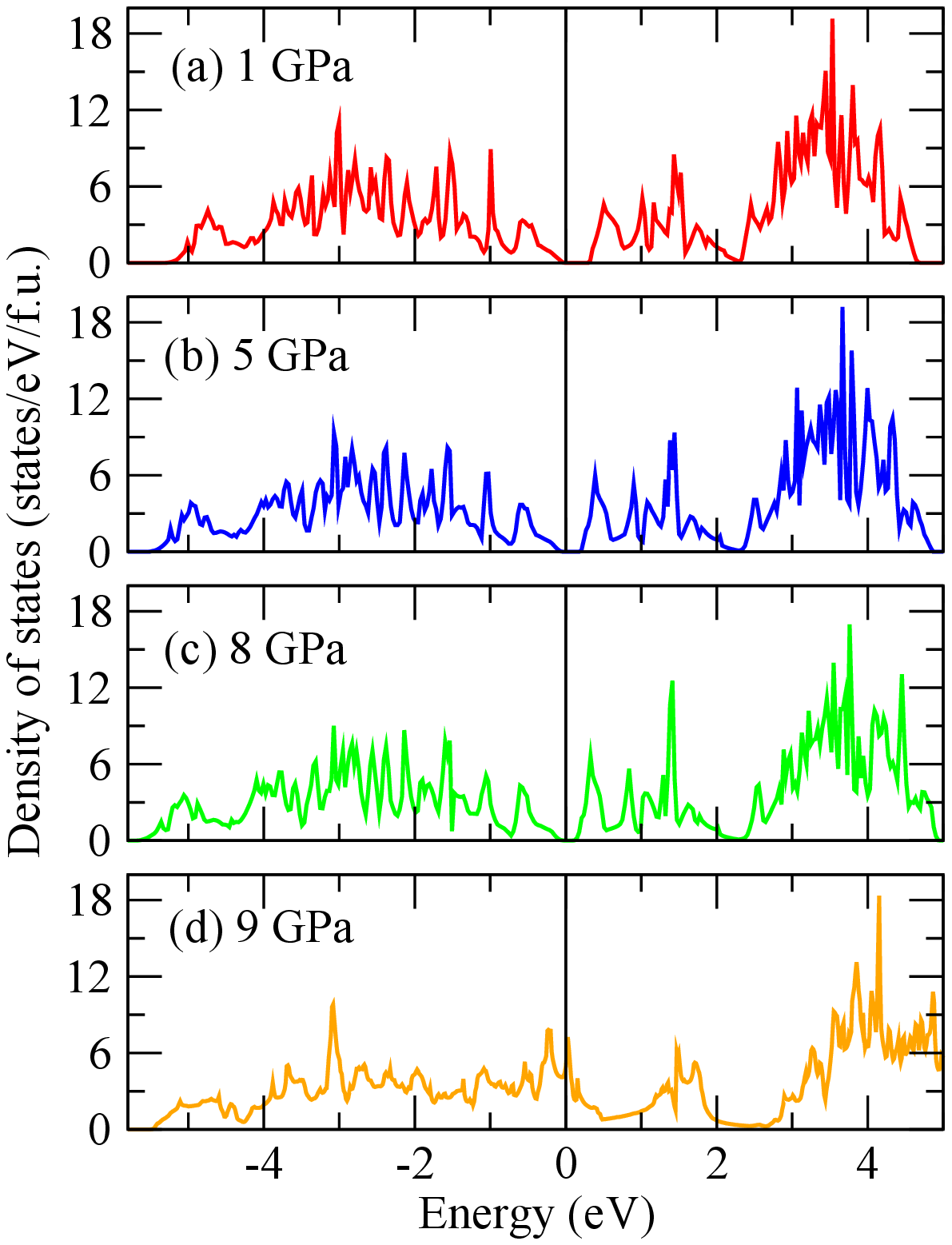}
		\caption{Calculated density of states of ThMnAsN with C-AFM spin configuration at (a) 1 GPa, (b) 5 GPa, (c) 8 GPa and (d) 9 GPa pressure. Fermi level is denoted by a vertical black line at $E=0$ eV. }
		\label{dosP}
	\end{figure}
	
		\begin{figure}[h!]
		\centering
		\includegraphics [height=6.0cm,width=8.5cm]{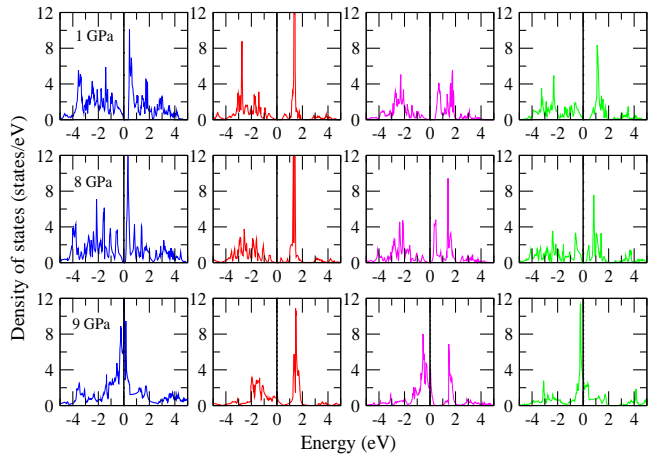}
		\caption{Mn-d orbital projected density of states of ThMnAsN at various pressure. The different color codes bear the same significance as those in Fig. \ref{cafmbs}. Orbital selective band gap is the special feature worth noticing.}
		\label{Odos}
	\end{figure}
Having presented in detail the effect of hydrostatic pressures on the structural parameters and associated modifications in elastic constants, it is imperative to present the effects of such changes of structural parameters on electronic band structure, density of states and Fermi surface at ambient as well as at higher pressure in this subsection. As mentioned earlier, in order to create C-AFM spin configuration a supercell $\sqrt 2 \times \sqrt 2 \times 2 $ structure is used. With a larger real-space cell, the reciprocal-space Brillouin Zone (BZ) will become smaller. As a result, the band structure will fold into the small supercell BZ. Electronic eigen states calculated for the supercell can be unfolded into the reciprocal space of the primitive cell directly in VASP, these technicalities are described in \cite{Dirnberger2021}. The electronic (as well as orbital projected) band structures at ambient and higher pressures are presented in Figs. \ref{cafmbs}, \ref{5cafmbs}, \ref{band} and \ref{orb} respectively. We depict the unfolded band structure in Fig.\ref{cafmbs} for ThMnAsN with C-AFM spin configuration at ambient pressure. A band gap of $\sim$ 0.4 eV is observed in ThMnAsN at ambient pressure and it can be compared with the experimentally measured band gap of 1.2 eV for LaMnAsO, a compound iso-structural to ThMnAsN \cite{McGuire2016}. Contribution of various Mn-d orbitals in the band dispersions are presented in the bottom Fig.\ref{cafmbs}b. Mn d$_{xz+yz}$, d$_{xy}$, d$_{z^2}$ and d$_{x^2-y^2}$ orbitals are denoted by blue, red, magenta and green colours respectively. The Mn d$_{xz+yz}$ and d$_{x^2-y^2}$ derived bands approach the Fermi Level both from the conduction and valance band, reducing the band gap as the pressure is increased (cf Figs. \ref{cafmbs}, \ref{5cafmbs}, \ref{band}). At 9 GPa pressure, there is no band gap and the system is metallic.
However a closer look at different orbital selective bands in Figs. \ref{cafmbs}, \ref{5cafmbs}, suggest orbital selective doping; the green band i.e, d$_{x^2-y^2}$ band gets influenced similar to as being hole doped whereas the magenta coloured band i.e, d$_{z^2}$ derived band near $\Gamma$ point suffers the effect of electron doping. This is a precursor to orbital selective Lifshitz transition usually observed as well as predicted in Fe-based superconductors \cite{MingY2017, Sen2020PRM, Ghosh2020, Ghosh2021, Ghosh2022, Craco2021}.
\par In Fig.\ref{band}, we depict our calculated band structure of non magnetic ThMnAsN at 9 GPa. Positions of Fermi level at different chemical potentials are also indicated in the band structure with a dotted line which is used to mimic the electron doping in the non magnetic ThMnAsN system within the rigid band model. Visibly different nature of electronic band structure above the magneto-structural transition is worth noticing. With the value of chemical potential $\mu$=0.25, we find a electronic band dispersion very similar to that of the generic band dispersion of Fe based superconductors. We also find that orbital nature of the low energy bands of electron doped ThMnAsN (within rigid band approximation with $\mu=0.25$) at 9 GPa pressure are also quite similar to the low energy band dispersion of most of the Fe based superconductors (see Fig.\ref{orb}). Moreover, domination of Mn d$_{xz+yz}$ and d$_{x^2-y^2}$ orbitals at the Fermi level at $\Gamma$ and M points which is quite similar to the band structure of stochiometric ThFeAsN \cite{2Wang2016, Singh2016, Sen2020PRB}, a Fe based superconductor iso-structural to ThMnAsN.
 
 \par Our calculated density of states at different hydrostatic pressures are presented in Fig.\ref{dosP}. We can clearly see, with the increasing pressure metallic behaviour is increasing as the energy gap is decreasing and at 9 GPa pressure, no energy gap exists at the Fermi level (cf. Fig. \ref{dosP}). This suggests that ThMnAsN is at the threshold of a metal insulator transition which is triggered by the collapse of local Mn moment and this transition may occur with the variation of pressure, temperature or may be even with the doping. We found a contrasting behaviour in LaMnPO, a iso-structural compound to ThMnAsN, where AFM to PM and insulator to metallic phase transitions occur at different pressures (not simultaneous). In LaMnPO, insulating AFM to metallic AFM transition occurs at 20 GPa pressure and metallic AFM to PM transition occurs at 34 GPa pressure \cite{Guo2013}. However, in this work, we have not studied the effect of doping and temperature in ThMnAsN. Nevertheless, we find three possible phase transitions (AFM to PM, tetragonal to collapse tetragonal and insulator to metal) that take place at around 9 GPa pressure. We also depict our calculated Mn-d orbital projected density of states for ThMnAsN at different hydrostatic pressure in Fig.\ref{Odos}. The orbital selective band gaps (cf. Fig. \ref{Odos}) at finite hydrostatic pressures are one of the important special features of ThMnAsN (this suggests possibility of orbital selective optics up to 8 Gpa). In this context, it is worth mentioning that the orbital selective modulation of electronic structure is also observed in ThFeAsN \cite{Sen2020PRM, Craco2021}.
\begin{figure}[h!]
		\centering
		\includegraphics [height=4.25cm,width=8.8cm]{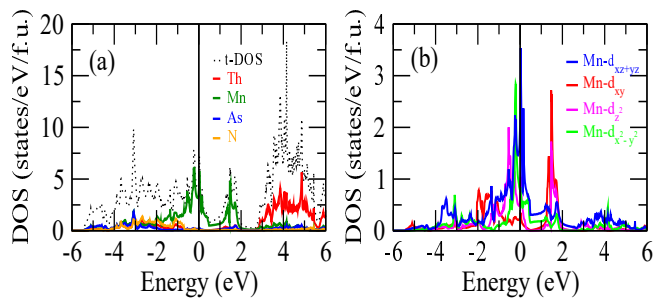}
		\caption{Calculated (a) atom projected and (b) Mn-3d orbital projected density of states of non-magnetic ThMnAsN at 9 GPa pressure.}
		\label{9dos}
	\end{figure}
	
	At 9 GPa pressure, atom projected density of states suggest that among all other atoms in ThMnAsN, Mn atoms have mostly contributed to the total density of states around the Fermi level (see Fig.\ref{9dos}a). Therefore, Mn d-orbital projected density of states for non magnetic ThMnAsN at 9 GPa pressure is presented in Fig.\ref{9dos}b. All five Mn-d orbitals are present at the Fermi level. Among them, Mn d$_{xz+yz}$ and d$_{x^2-y^2}$ orbitals dominate the Fermi level.  
	\begin{figure}[h!]
		\centering
		\includegraphics [height=6.5cm,width=8cm]{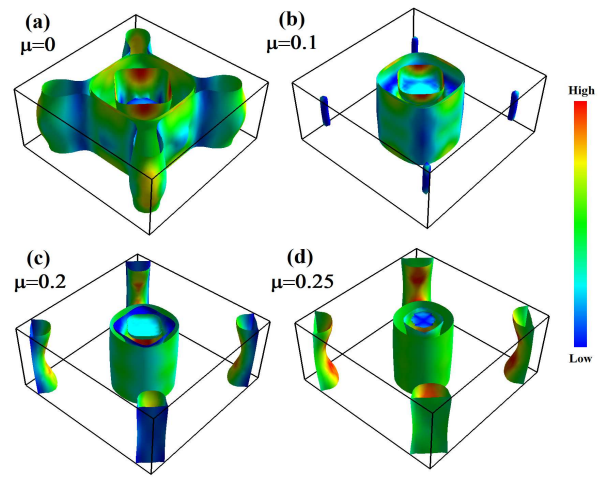}
		\caption{Calculated Fermi surfaces of non-magnetic ThMnAsN at 9 GPa pressure shaded by band velocity for (a) $\mu=0$, (b) $\mu=0.1$, (c) $\mu=0.2$ and (d) $\mu=0.25$.}
		\label{fs}
	\end{figure}
 \begin{figure}[h!]
		\centering
		\includegraphics [height=7.5cm,width=7.5cm]{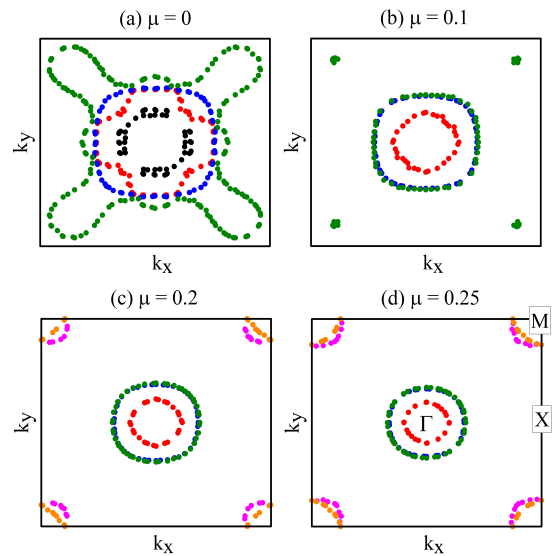}
		\caption{Calculated Fermi surfaces section of non-magnetic ThMnAsN in 001 plane at k$_z$=0 at 9 GPa pressure for (a) $\mu=0$, (b) $\mu=0.1$, (c) $\mu=0.2$ and (d) $\mu=0.25$.}
		\label{FScut}
	\end{figure}
 \begin{figure}[h!]
		\centering
		\includegraphics [height=4.0cm,width=7.5cm]{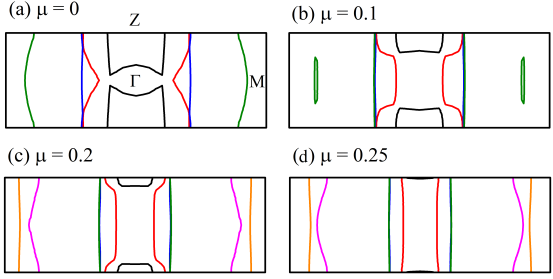}
		\caption{Calculated Fermi surfaces section of non-magnetic ThMnAsN in 110 plane, showing k$_z$ dispersion of Fermi surfaces at 9 GPa pressure for (a) $\mu=0$, (b) $\mu=0.1$, (c) $\mu=0.2$ and (d) $\mu=0.25$.}
		\label{FSz}
	\end{figure}
To illustrate the metallic behaviour of ThMnAsN at 9 GPa pressure more closely, we calculate the FSs of ThMnAsN. Fermi surfaces of ThMnAsN at 9 GPa pressure in the non magnetic phase for different chemical potentials which are depicted in Fig.\ref{fs}. Fermi velocities are also indicated in all the Fermi surfaces. Fermi surface topology, specially with $\mu=0.25$, closely resembles with the generic Fermi surface of the Fe based superconductor \cite{Sen2014, Sen2015, Paglione2010, Sen2017s}. It contains three hole like Fermi surfaces in the centre of the Brillouin zone ($\Gamma$ point) and two electron like Fermi surfaces at the corner of the Brillouin zone (M point). In Fig.\ref{FScut} and \ref{FSz}, we display the Fermi surface sections of ThMnAsN in 100 and 110 (k$_z$ dispersion) plane respectively. From Fig.\ref{FScut} and \ref{FSz}, it is quite evident that with the higher values of chemical potential the Fermi surfaces become more 2 dimensional (2D) in nature. Appearance and disappearance of various Fermi surfaces in the BZ with the variation of chemical potential (introduced to mimic the effect of electron doping) bears the signature of orbital selective Lifshitz transition. 2D Nature of these Fermi surfaces are also favourable for nesting (with electron doping) which plays a very important role in pnictides superconductors to give rise to various exotic phases like spin density wave order (SDW), orbital density wave order (ODW) {\it etc.,} including superconductivity \cite{Terashima2019, Sen2015, Ghosh2017, 1Sen2015, Sunagawa2015}. From Figs. \ref{fs}, \ref{FScut} and \ref{FSz}, it is clear that there is formation of electron type cylinders at around $\Gamma$ point whereas hole type cylindrical FSs at around X point. Strong reconstruction in electronic structure (and hence transport properties) similar to KFe$_2$As$_2$ \cite{Stavrou2015}; with slight electron doping, electron like charge carriers are missing around M point whereas with further electron doping electron FSs again grow around the BZ corners (M points) and there is possibility of nesting between the inner hole to electron FSs.  Since Fermi surface topology and low energy electronic structure are believed to be the main controlling factors of superconductivity in Fe based superconductors, ThMnAsN at high pressure has a great potential to become a high temperature superconductor with unconventional superconducting properties.
 
 \begin{figure}[h!]
		\centering
		\includegraphics [height=6.5cm,width=7.5cm]{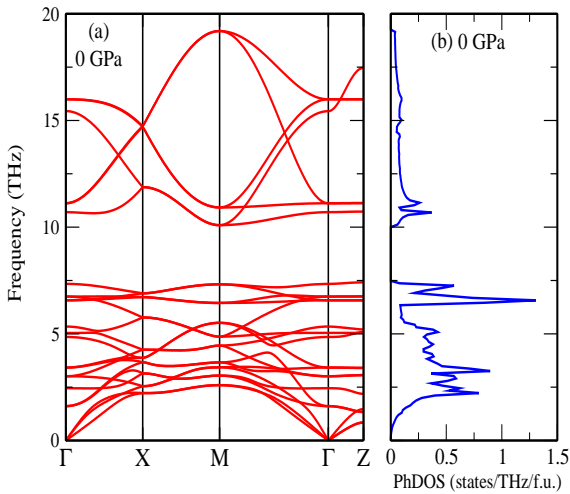}
		\caption{(a) Phonon dispersion and (b) phonon density of states of ThMnAsN at ambient pressure with C-AFM spin configuration.}
		\label{TMANph0}
	\end{figure}
		\begin{figure}[h!]
		\centering
		\includegraphics [height=6.5cm,width=7.5cm]{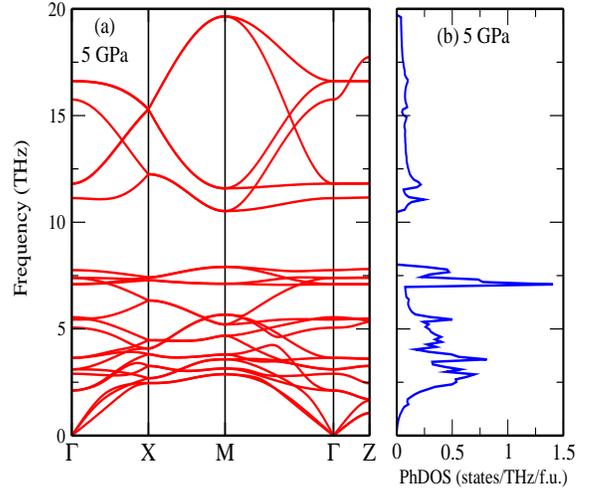}
		\caption{(a) Phonon dispersion and (b) phonon density of states of ThMnAsN at 5 GPa pressure with C-AFM spin configuration.}
		\label{TMANph5}
	\end{figure}
		\begin{figure}[h!]
		\centering
		\includegraphics [height=6.5cm,width=7.5cm]{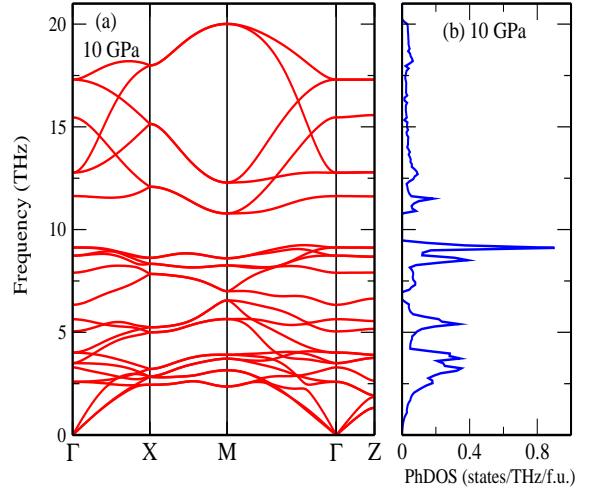}
		\caption{(a) Phonon dispersion and (b) phonon density of states of non-magnetic ThMnAsN at 10 GPa pressure.}
		\label{TMANph10}
	\end{figure}
 
   To probe the dynamical stability of ThMnAsN at various pressure we perform phonon calculations. In Fig.\ref{TMANph0}, \ref{TMANph5} and \ref{TMANph10}, we present our calculated phonon dispersions and phonon density of states at 0, 5 and 10 GPa pressures respectively for ThMnAsN. For phonon calculations at ambient and 5 GPa pressure, we use C-AFM spin configuration for ThMnAsN. On the other hand, for ThMnAsN at 10 GPa pressure, we performed non-magnetic phonon calculation. Absence of any imaginary frequencies clearly appraise the dynamical stability of the system at various pressures. The gap between the optical and the acoustic phonon branch in phonon density of states decreases with pressure that has important bearings on the phonon thermal conductivity and hot carrier relaxation processes. The phonon dispersion becomes flatter in $\Gamma - Z$ direction at higher pressures. The phonon density of states decreases in the low phonon frequencies at high pressures. 
  
\section{Conclusions}
A detailed density functional theory based first principles studies are carried out on ThMnAsN. Our study predicts simultaneous C-AFM tetragonal to paramagnetic collapse tetragonal (PM -cT) transition when the Mn moments are completely quenched at the collapse tetragonal phase. All the structural parameters bear the finger prints of (PM -cT) phase transition at around 8 to 9 GPa pressure which is attributed to a iso-structural phase transition from a tetragonal to a collapse tetragonal phase. A comparison with the earlier discovered 122 family of Fe-based superconductors (like Ca(K/Na)Fe$_2$As$_2$ \cite{Kreyssig2008, Stavrou2015, Long2013, Uhoya2010, Mittal2011, Nakajima2015}) where similar transitions were observed, indicates similar mechanism of such transition (but not the same) in ThMnAsN but no such transition was predicted/discovered in 1111-type Fe-based superconducting materials so far. In contrast to the earlier observed tetragonal to collapse tetragonal phase transition in 122 family of Fe-based superconductors \cite{Kreyssig2008, Stavrou2015, Long2013, Uhoya2010, Mittal2011, Nakajima2015} there is no direct covalent bond formation between the inter-planer As-As but as the out of plane As-As distance becomes less than the in-plane As-As bond length, all 4-As atoms strongly overlap with the Mn atom. This leads to indirect overlapping between the Mn atoms via As atoms (see Fig. \ref{isop}). This destroys the otherwise "localized" nature of Mn moments and moments get completely quenched above P$_c$.
\par We also evaluate the variation of various elastic constants of tetragonal ThMnAsN with pressure. We show that with the increasing pressure, metallic behaviour of the system enhances which is supported by our pressure dependent density of states, charge density plots and COHP calculations. Moreover, we have seen a significant increase of positive Cauchy pressure C$_{13}$-C$_{44}$ around 9 GPa pressure, also indicating an abrupt uplift of the metallic character in ThMnAsN. Influence of hydrostatic pressure on electronic band structure, Fermi surface, density of states are also described in details. Dynamical stability of ThMnAsN was thoroughly verified at all presented calculations at various pressures by checking positive phonon frequency in the phonon dispersion curves. Phonon density of states exhibit that the phonon band gap (between optical and acoustic phonon branch) decreases with the increase of hydrostatic pressure.  
\par In general, electronic structure of ThMnAsN at high pressure resembles with the electronic structure of Fe-based superconductors with similar special features in common like orbital selectivity, Lifshitz transition etc. Therefore, a possible unconventional high temperature superconductivity may appear in ThMnAsN which might help to resolve the unsettled issues of Fe based superconductivity or in general high temperature superconductivity. An example of 
simultaneous structural, magnetic and metal-insulator transitions at the vicinity of 9 GPa pressure is found in ThMnAsN. We believe our work will stimulate further experimental and theoretical research in this direction.

\end{document}